**Lyudmyla Potrashkova,**  
*Ph.D., Associate professor,*  
*Semen Kuznets Kharkiv National University of Economics,*  
*Kharkiv, Ukraine*

**Diana Raiko,**  
*D.Sc., Professor, National Technical University «Kharkiv Polytechnic Institute»,*  
*Kharkiv, Ukraine*

**Leonid Tseitlin,**  
*Ph.D., National Technical University "Kharkiv Polytechnic Institute",*  
*Kharkiv, Ukraine*

**Olga Savchenko,**  
*Ph.D., Professor, National Technical University "Kharkiv Polytechnic Iinstitute",*  
*Kharkiv, Ukraine*

**Szabolcs Nagy,**  
*Ph.D., Associate Professor,*  
*Head of Department of Marketing Strategy & Communication,*  
*Faculty of Economics, Marketing & Tourism Institute, University of Miskolc,*  
*Miskolc, Hungary*

# Methodological provisions for conducting empirical research of the availability and implementation of the consumers' socially responsible intentions

*Social responsibility of consumers is one of the main conditions for the recoupment of enterprises' expenses associated with the implementation of social and ethical marketing tasks. Therefore, the enterprises, which plan to act on terms of social and ethical marketing, should monitor the social responsibility of consumers in the relevant markets. At the same time, special attention should be paid to the analysis of factors that prevent consumers from implementing their socially responsible intentions in the regions with a low level of social activity of consumers. The purpose of the article is to develop methodological guidelines that determine the tasks and directions of conducting empirical studies aimed at assessing the gap between the socially responsible intentions of consumers and the actual implementation of these intentions, as well as to identify the causes of this gap. An empirical survey of the sampled consumers in Kharkiv was carried out in terms of the proposed methodological provisions. It revealed a rather high level of respondents' willingness to support socially responsible enterprises and a rather low level of implementation of these intentions due to the lack of consumers' awareness. To test the proposed methodological guidelines, an empirical study of the consumers' social responsibility was conducted in 2017 on a sample of students and professors of the Semen Kuznets Kharkiv National University of Economics (120 people). Questioning of the respondents was carried out using the Google Forms. The finding allowed to make conclusion for existence of a high level of respondents' willingness to support socially responsible and socially active enterprises. However, the study also revealed the existence of a significant gap between the intentions and actions of consumers, caused by the lack of awareness. The authors allocated the prospects of socially responsible business development in the region, provided the active measures were taken by business and authorities to inform the consumers about the social responsibility and social activity of producers.*

**Introduction.** The processes of the post-industrial economy development and the spread of the sustainable development ideas have increased the requirements of society to the enterprises' activities: today, the society expects enterprises to adhere to the principles of social responsibility.

The concept of social and ethical marketing is the projection of ideas of the business social responsibility on marketing. Social and ethical marketing expands the set of criteria for companies to assess their marketing decisions: two traditional criteria of meeting the needs of consumers and obtaining the business benefits are supplemented by the criterion of considering the social interests. In other words, the manufacturers have to find a compromise between their own interests, the interests of specific consumers and the long-term interests of the entire society.

What matter is that the enterprise's implementation of the principles of social and ethical marketing may be useful for the enterprise. Thus, P. Kotler states that "some companies have achieved significant sales and profits by adopting and implementing the concept of social and ethical marketing" [1]. However, for social and ethical marketing to be business-beneficial, the society should be prepared for it. One of the main conditions for the recoupment of enterprises' expenses associated with the implementation of social and ethical marketing is the social responsibility of the consumers of the enterprise's products.

*The social responsibility of consumers* can be defined as the conscious willingness of the consumers to evaluate their choice of a product in terms of its impact on the society.

Although an increasing number of people share the principles of corporate social responsibility today, the number of socially responsible consumers cause certain doubts and require statistical review. Therefore, one of the stages of substantiating the expediency of the enterprise's activity in terms of social and ethical marketing should be the assessment and analysis of the consumers' social responsibility in the relevant market.

**Analysis of recent research.** Comprehension of the fact that the consumers' social responsibility is the driving force behind the development of a socially responsible business, has led to many empirical studies of the subject. The purpose of these studies is to identify the extent to which the consumers making purchasing decisions consider the characteristics of products and their manufacturers, which are important from the point of view of the interests of social sustainable development. The main tool for studying the social responsibility of consumers is the method of questioning, supplemented by the methods of mathematical statistics and econometrics.

Thus, a leading American sociological marketing company Nielsen has repeatedly conducted statistical surveys of various aspects of social responsibility of producers and consumers using the questionnaire method. In 2015, Nielsen conducted the Nielsen Global Survey of Corporate Social Responsibility and Sustainability, having interviewed online more than 30,000 consumers from 60 countries worldwide. The results of the carried out research [2] indicate that consumers in Ukraine and around the world consider certain characteristics of the social responsibility of producers of these products when making decisions to purchase food products, toiletries, and medicines (Table 1).

Using the methods of mathematical statistics and econometrics in addition to the method of questioning can solve the problem of identifying certain statistical dependencies, as well as certain discrepancies in relation to the corporate social responsibility from various population groups. For example, [4] uses the multiple regression model to compare the impact on customer loyalty in the US hotel industry of four factors, such as corporate social responsibility, corporate ability, reputation and transparency of the hotels. According to the research findings, the corporate social responsibility initiatives have the greatest impact on the customers' loyalty. In the same work, the Student's t-criterion revealed the discrepancies in relation to the corporate social responsibility of the hotels of men and women: women were more socially responsible consumers. [5] analyses the influence of the characteristics of producers' social responsibility on consumer judgments at various stages of the process of making consumer decisions in the cosmetics sector using the descriptive and multidimensional statistical analysis. [6] deals with the consumers' attitudes toward various product characteristics and their manufacturers in the footwear market using the conjoint analysis. The research has shown that the producers' environmental responsibility significantly affects the consumers' purchasing decision.

*Table 1* – **Characteristics of the producers' social responsibility affecting the consumers' purchasing decision, according to** (2, 3)

| № | Characteristics of food products, toiletries and medicines in terms of the producers' social responsibility | Percentage of the respondents that considered these characteristics when making their purchasing decisions* | |
|---|---|---|---|
| | | *Worldwide* | *In Ukraine* |
| 1 | The products are manufactured by the brand/company I trust | 62 | 54 |
| 2 | The products are beneficial to health | 59 | 61 |
| 3 | The products are produced by the company which cares about the environment | 45 | 29 |
| 4 | The products are produced by the socially responsible company | 43 | 29 |
| 5 | The product has environment-friendly packaging | 41 | 27 |
| 6 | The products are produced by the company, which invests in the development of the local community | 41 | 24 |

\* Percentage of the respondents who rated the factor as having "very strong influence" or "strong influence" on their purchasing decision the week prior to the survey

The general conclusion of the empirical research of the consumers' social responsibility is optimistic: the producers' social responsibility positively influences their consumers' purchasing decision.

**Selection of the previously unresolved issues.** Empirical studies [6-9] have shown that the degree of influence of the characteristics of producers' social responsibility on their consumers' purchasing decisions varies from one country to another. Therefore, the analytical work needs to be continued and supplemented with the domestic empirical material. At the same time, due to the specifics of domestic realities, particular attention should be paid to the analysis of factors that prevent Ukrainian consumers from implementing their socially responsible intentions. Previous researchers did not pay enough attention to the analysis of such factors; therefore, it has determined the purpose of our study.

**The purpose of the research.** The purpose of the article is to develop methodological guidelines to determine the tasks and directions of conducting empirical studies aimed at assessing the gap between the consumers' socially responsible intentions and the implementation of these intentions, as well as to identify the causes of this gap.

**Methodological provisions on conducting empirical studies on the availability and implementation of consumers' socially responsible intentions.** The decision of the company's management to act in accordance with the principles of social and ethical marketing needs justification based on the assessment of the economic efficiency of such activities. One should consider that the effectiveness of social and ethical marketing depends largely on the level of the consumers' social responsibility in the relevant market. Thus, the following analytical tasks facing the management of the enterprise emerge:

1. Evaluating the cost-effectiveness of implementing the principles of social and ethical marketing at the enterprise - taking into account the level of the consumers' social responsibility (Fig. 1).

2. Implementation of this task requires the conducting of empirical studies aimed at determining the level of the consumers' social responsibility in the relevant market.

3. Planning of measures aimed at increasing the enterprise economic efficiency in terms of social

and ethical marketing by increasing the consumers' social responsibility and removing the obstacles to its implementation.

Fulfillment of this task requires the conducting of empirical research aimed at identifying the obstacles for the implementation of the consumers' socially responsible intentions in the relevant market.

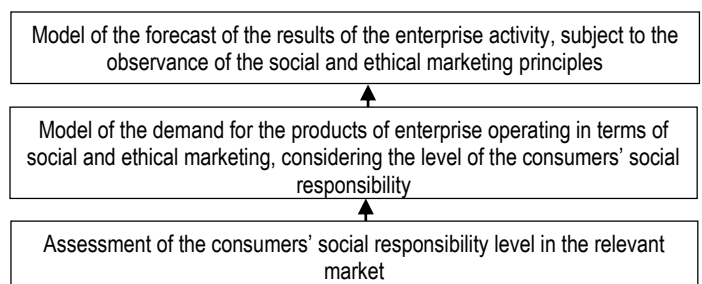

Figure 1 – **The main components of the mathematical toolset for assessing the economic effects of implementation of the principles of social and ethical marketing, developed by the author**

For empirical studies of the consumers' social responsibility to contribute to the solution of the stated tasks of the enterprise management, they must meet the following target requirements:

1. The task of studying the consumers' social responsibility:

a. identification of the consumers' intentions in the relevant market as to the consumers' choice of the socially responsible and socially active enterprises;

b. determination of the level of actual implementation of the consumers' socially responsible intentions;

c. identification of the reasons that prevent consumers from implementing their socially responsible intentions;

2. The method of obtaining information is the questioning;

3. Use of the received information:

a. formation of the measures of enterprises and state authorities aimed at eliminating the obstacles to the implementation of the consumers' socially responsible intentions;

b. substantiation of the decisions as to the implementation of the social and ethical marketing principles at the enterprises.

We will describe the areas to carry out the empirical study of the consumers' social responsibility in order to facilitate the resolution of the above-mentioned tasks of the enterprise management.

Area 1. Analysis of the actual consumers' behavior. This area of research is aimed at identifying the extent to which the consumers consider the characteristics of products and their producers that are important from the point of view of the social interests (its sustainable development) when making purchasing decisions. Such characteristics are as follows: conformity of the product and its packaging to the environmental safety requirements; compliance of the production process with animal protection requirements; compliance of the manufacturer activities with the principles of environmental liability and liability to employees and consumers; manufacturer's charity activities and involvement in the socially significant projects.

Within this area of research, the respondents are offered a list of product characteristics that are important from the point of view of the social interests, and each respondent is asked to mark those characteristics he usually considers when making purchasing decisions.

According to the results of the questionnaire, each characteristic of the products from a given list is considered by the percentage of consumers when actually making their consumer decisions. We denote this percentage as $\alpha_i$, where *i* is the number of the corresponding product characteristics in terms of social responsibility, *i=1,…,N*. Obtained values $\{\alpha_i \mid i=1,…,N\}$ will allow us to draw conclusions about what product characteristics are important from the point of view of the social interests, largely affect the actual decisions of consumers in the purchasing process, and as a result, what product characteristics should be improved by the manufacturers to enhance their reputation

Area 2. Analysis of the consumers' intentions. This area of research is aimed at identifying the consumers' intentions to consider the products' characteristics important from the point of view of the social interests when making purchasing decisions.

The formation of the methodological foundations of this area of empirical research is based on the theory of the planned behavior of I. Eisen [10], according to which the behavior of people is determined by their intentions and the ability to implement these intentions.

One can assume that implementation of the consumers' socially responsible behavior depends on such factors as:
− consumers' socially responsible intentions;
− consumers' awareness about the social responsibility and social activity of enterprises;
− consumers' ability to choose the products and to give up the products made by irresponsible producers (Fig. 2).

One should also remember about the availability of the "socially responsible consumption funnel ". At the entrance to this funnel, there is a large number of people with the intention of socially responsible consumption. However, the proportion of these people is shrinking because of the lack of information about the environmental friendliness of the products and the behavior of producers. Further, a part of those remaining is shattered because of the lack of opportunities for replacing the products made by irresponsible producers. As a result, socially responsible consumption is implemented only by a part of responsible consumers.

The consumers' intentions to implement their social responsibility when purchasing the products can be assessed by obtaining such information from the respondents' questionnaires:
− determining the percentage of consumers willing to pay more for the environmentally safe products;
− determining the percentage of consumers who support environmental consumption, which is expressed in the conscious choice of goods and services produced with minimal harm to the environment;
− determining the percentage of consumers who are willing to pay more for the products of charitable enterprises;
− determining the percentage of consumers who agree that charitable enterprises and socially significant projects deserve greater confidence among the customers than other enterprises;
− determining each given characteristic of the products, which is important from the point of view of the social interests, the percentage of consumers who would like to consider this characteristic when making purchasing decisions.

Denote $\beta_i$ as the percentage of consumers who would like to consider the *i*-th characteristic of the products, important from the point of view of the social interests, when making their consumer decisions. Then the difference between the values $\beta_i$ and $\alpha_i$ shows the part of consumers who did not implement their socially responsible intentions as for the consideration of the relevant product characteristics. The information obtained will allow us to draw conclusions about what socially responsible intentions of the consumers are actually implemented most of all as well as which ones remain unimplemented.

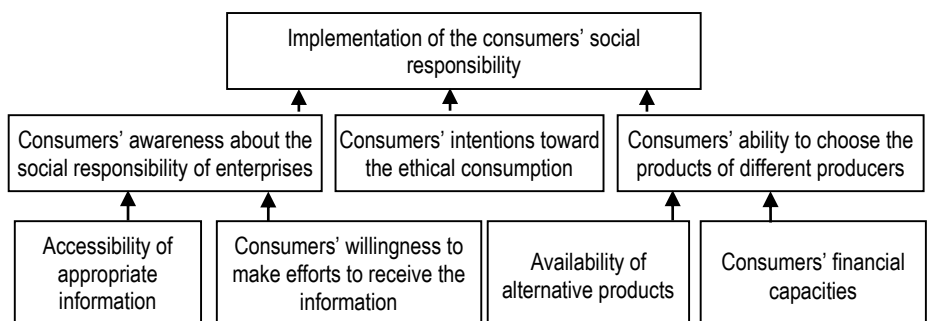

Figure 2 – **The scheme of factors that affect the implementation of consumers' social responsibility** (developed by the author)

Area 3. Analysis of the reasons that prevent consumers from implementing their socially responsible intentions. This area of research is aimed at identifying the main obstacles that prevent the implementation of the consumers' socially responsible intentions. A preliminary list of such obstacles is expedient to include:

1. lack of information:
a. lack of general information about the characteristics of products and their producers that are important in terms of sustainable development of the society;
b. lack of information about the social responsibility and social activity of the specific products' manufactures;
2. lack of consumer choice of products, as well as the refusal to buy the products made by irresponsible producers:
a. as a result of the absence of alternative product options;
b. due to the low financial capacities of consumers.

Conclusions on the existing barriers to the implementation of the consumers' socially responsible intentions can be made by obtaining such information from the respondents' questionnaires:

− determining the percentage of consumers who are not familiar with the idea of social responsibility of business;

− determining the percentage of consumers who believe that their lack of information hampers the implementation of their social responsibility;

− determining the percentage of consumers who believe that their social responsibility is hindered by their insufficient income;

− determining the percentage of consumers who believe that their social responsibility is hampered by lack of time;

− determining the percentage of consumers who believe that they lack information on the environmental and social responsibility of the products' manufactures;

− determining the percentage of consumers who visited the producers' sites to assess their environmental and social responsibility;

− determining the percentage of consumers who pay attention to the environmental marking of their products.

The obtained information will allow drawing conclusions about the main obstacles to the implementation of the consumers' socially responsible intentions, as well as on possible measures taken by the enterprises and authorities focused on eliminating these obstacles.

Area 4. Defining the gaps in intentions and implementation of the social responsibility of different groups of consumers. This research area aims to define the differences in attitude toward the corporate social responsibility of men and women, people of all ages and other consumer groups.

The obtained information will allow specifying the parameters of measures aimed at eliminating the obstacles to implementation of the consumers' social responsibility, considering the characteristics of different population groups.

**Empirical research of the consumers' social responsibility.** To test the proposed methodological guidelines, an empirical study of the consumers' social responsibility was conducted in 2017 on a sample of students and professors of the Semen Kuznets Kharkiv National University of Economics (120 people). Questioning of the respondents was carried out using the Google Forms.

The research has revealed the following facts:

− the respondents consider certain characteristics of the social responsibility of producers when making their purchasing decisions on consumer goods (Table 2, column 3);

− there is a significant gap between the consumers' socially responsible intentions and the implementation of these intentions (Table 2, Columns 2 and 3).

− the respondents consider the lack of information the main obstacle that prevents consumers from implementing their socially responsible intentions today (Table 3).

*Table 2* – **Consumers' intentions and actual consideration when choosing the product with the characteristics of the producers' social responsibility** (based on the empirical research)

| Characteristics of the producers' social responsibility | Percentage of the respondents intending to consider these characteristics when making purchase, $\beta_i$ | Percentage of the respondents actually considering these characteristics when making purchase, $\alpha_i$ |
|---|---|---|
| *Positive characteristics* | | |
| The products are manufactured by the socially responsible company | 61,2 % | 37,6 % |
| The products and their packaging are not harmful to the environment | 77,6 % | 30,6 % |
| The products were not tested on animals (household detergents and makeup) | 51,8 % | 27,1 % |
| The products are manufactured by the charitable enterprise | 25,9 % | 11,8 % |
| A part of the proceeds from the sale of products will go to charity | 41,2 % | 16,5 % |
| *Negative characteristics* | | |
| The products are manufactured by the enterprise that violates the consumers' rights and produces low-quality products | 90,6 % | 81,2 % |
| The products are manufactured by the enterprise that pollutes the environment violating the environmental standards | 76,5 % | 35,3 % |
| The products are manufactured by the enterprise that violates the employees' rights | 40,0 % | 11,8 % |

A direct question was: "Do you have enough information to judge about the environmental and social responsibility of the enterprises manufacturing the products you buy?". 92% of the respondents replied that they did not have enough information, 7% of the respondents replied that the information "is rather sufficient". At the same time, only 6% of the respondents repeatedly visited the business sites to find the right information, and only 31% of the respondents often or always pay attention to the environmental marking of the products.

*Table 3* – **The main reasons that hinder consumers from implementing their socially responsible intentions** (based on the empirical research)

| Factors that prevent consumers from implementing their socially responsible intentions (according to the respondents) | Percentage of the respondents |
|---|---|
| Lack of information about the environmental and social responsibility of the enterprises | 73 % |
| Insufficient income | 27 % |
| Lack of time | 20 % |

The content of the concept of "social responsibility of business (enterprises)" is comprehensive for 75% of the respondents.

The following figures indicate the intention of the respondents to support environmentally responsible and socially active behavior of enterprises:

84% of the respondents support the idea of ecological consumption, and 17% are trying to practice it; almost all respondents (95%) are willing to pay a little more for the environmentally safe products;

about half of the respondents (47%) trust the companies that carry out socially significant projects more than others; 59% of the respondents are willing to pay more for the products of charity businesses.

The analysis of the differences between different sex samples showed interesting results. There were statistically significant differences in the level of socially responsible intentions of men and women: women would like to consider more social product characteristics than men (on average, women chose 5-6 characteristics from the list, while men choose about 4). At the same time, a greater proportion of women lack the information about the social responsibility of producers. However, in fact, women and men consider approximately the same number of social characteristics of the products - about three out of the above-given list.

**Conclusions.** Social responsibility of consumers is one of the main conditions for the recoupment of enterprises' expenses associated with the implementation of social and ethical marketing tasks. Therefore, the enterprises, which plan to act on terms of social and ethical marketing, should monitor the social responsibility of consumers in the relevant markets. Moreover, enterprises must identify the possibility of their influence on the conditions for the implementation of the consumers' social responsibility. Fulfillment of these tasks requires empirical studies of the consumers' social responsibility in the region. In order to develop the theoretical foundations of such studies, the article proposes methodological guidelines for the assessment and identification of the causes of the gap between the consumers' socially responsible intentions and their implementation.

The novelty of the proposed methodological provisions is determined by the application of a systematic approach to the analysis of the consumers' social responsibility, according to which a system of factors is considered to fulfill the purposes of the analysis. The factors are as follows: consumers 'socially responsible intentions; a set of obstacles that hinder the realization of these intentions; characteristics of actual consumer behavior in terms of social responsibility.

The practical value of the proposed methodological provisions is determined by their intended use for providing the information support to the management of enterprises in terms of social and ethical marketing.

An empirical survey of the sampled consumers in Kharkiv showed a high level of respondents' willingness to support socially responsible and socially active enterprises. However, the study also revealed the existence of a significant gap between the intentions and actions of consumers, caused by the lack of awareness. The obtained results testify to the prospects of development of socially responsible business in the region, provided the active measures are taken by business and authorities to inform the consumers about the social responsibility and social activity of producers.

The proposed methodological provisions require further development in the area of analysis of the processes of formation and implementation of the social responsibility of different consumer groups.

***Л. В. Потрашкова,*** к. е. н., доцент, Харківський національний економічний університет імені Семена Кузнеця, (Харків, Україна);

***Д. В. Райко,*** д. е. н., професор, Національний технічний університет «Харківський політехнічний інститут» (Харків, Україна);

***Л. М. Цейтлін,*** к. ф.-м. н., докторант, Національний технічний університет «Харківський політехнічний інститут» (Харків, Україна);

***О. І. Савченко,*** к. е. н., доцент, Національний технічний університет «Харківський політехнічний інститут» (Харків, Україна);

***Н. Саболч,*** Ph.D., доцент, Університет Мішкольца (Мішкольц, Угорщина).

**Методологія проведення емпіричних досліджень наявності та реалізації соціально відповідальних намірів споживачів**

*Соціальна відповідальність споживачів є однією з основних умов окупності витрат підприємств на виконання завдань соціально-етичного маркетингу. З цього випливає, що підприємства, які планують діяти на засадах соціально-етичного маркетингу, мають здійснювати моніторинг соціальної відповідальності споживачів на відповідних ринках. При цьому у регіонах з невисоким рівнем соціальної активності споживачів, необхідно приділяти особливу увагу аналізу факторів, які перешкоджають споживачам реалізовувати їхні соціально відповідальні наміри. Метою статті є розробка методологічних положень, які визначають завдання та напрями проведення емпіричних досліджень, спрямованих на оцінювання розриву між соціально відповідальними намірами споживачів та фактичною реалізацією цих намірів, а також на виявлення причин цього розриву. Для апробації запропонованих методологічних положень у 2017 році було здійснено емпіричне дослідження соціальної відповідальності споживачів на прикладі вибірки студентів та викладачів Харківського національного економічного університету імені Семена Кузнеця (120 осіб). Анкетування респондентів здійснювалось за допомогою Google Forms. Отримані результати дозволили авторам прийти до висновку про наявність високого рівня готовності респондентів підтримувати соціально відповідальні та соціальні активні підприємства. Але при цьому дослідження виявило і наявність значного розриву між намірами та діями споживачів, спричиненого їхньою недостатньою інформованістю. Авторами виокремлено перспективи розвитку соціально відповідального бізнесу в регіоні при умові здійснення активних заходів з боку бізнесу та органів влади щодо інформування споживачів про соціальну відповідальність та соціальну активність виробників.*

Ключові слова: соціальна відповідальність споживачів, соціальна відповідальність бізнесу, соціально-етичний маркетинг, інформованість споживачів, метод анкетування.